\documentclass[journal=nalefd,manuscript=article,reprint, 
 amsmath,amssymb,
]{achemso}
\usepackage{mathrsfs}
\usepackage{color}
\usepackage{graphicx}
\usepackage{dcolumn}
\usepackage{bm}
\usepackage{amsmath,mathrsfs}    
\usepackage{graphicx}   
\usepackage{subfigure}  
\usepackage{hyperref}   
\usepackage{upgreek}
\usepackage{textcomp}
\usepackage[T1]{fontenc}

\title{Charge Detection in Gate-Defined Bilayer Graphene Quantum Dots}
\author{Annika~Kurzmann}
\email{annikak@phys.ethz.ch}
\author{Hiske~Overweg}
\author{Marius~Eich}
\author{Alessia~Pally}
\author{Peter~Rickhaus}
\author{Riccardo~Pisoni}
\author{Yongjin~Lee}
\affiliation{Solid State Physics Laboratory, ETH Zürich, CH-8093 Zürich, Switzerland}
\author{Kenji~Watanabe}
\author{Takachi~Taniguchi}
\affiliation{National Institute for Material Science, 1-1 Namiki, Tsukuba 305-0044, Japan}
\author{Thomas~Ihn}
\author{Klaus~Ensslin}
\affiliation{Solid State Physics Laboratory, ETH Zürich, CH-8093 Zürich, Switzerland}

\date{\today}
\begin{document}

\begin{abstract}
	We report on charge detection in electrostatically-defined quantum dot devices in bilayer graphene using an integrated charge detector. The device is fabricated without any etching and features a graphite back gate, leading to high quality quantum dots. The charge detector is based on a second quantum dot separated from the first dot by depletion underneath a 150 nm wide gate. We show that Coulomb resonances in the sensing dot are sensitive to individual charging events on the nearby quantum dot. The potential change due to single electron charging causes a step-like change (up to 77\%) in the current through the charge detector. Furthermore, the charging states of a quantum dot with tunable tunneling barriers and of coupled quantum dots can be detected. 
	\end{abstract}

\maketitle

\section{Introduction}
Graphene is a promising candidate for future nano-electronic devices including building blocks for quantum information processing. Reasons are the expected long spin lifetimes \cite{trauzettel2007spin} and high carrier mobilities \cite{novoselov2005two,zhang2005experimental,banszerus2015ultrahigh}. Experimentally these spin lifetimes have not been demonstrated yet. For progress in this direction \cite{elzerman2004single} a device is needed that allows to confine charges and simultaneously measure their dynamics in a time-resolved way. This is possible with a charge detector in close proximity to a graphene quantum dot (QD). In GaAs based devices a combination of QDs and QPCs allowed to detect spin-qubit states \cite{elzerman2004single,petta2005coherent} and molecular states in coupled QDs \cite{dicarlo2004differential}. Furthermore, charge detection can be used to investigate tunneling dynamics of charges in a time-resolved way \cite{gustavsson2006counting,vandersypen2004real} and to obtain the full counting statistics of the charge current and the charge occupation. 
 
Recent experiments show the fabrication and measurement of quantum dots \cite{eich2018spin,Ban,doi:10.1021/acs.nanolett.8b01859} and quantum point contacts \cite{overweg2017electrostatically,Ban,kraft2018valley} in bilayer graphene, that are comparable to GaAs devices. Bilayer graphene offers the possibility to electrostatically define nanostructures by opening a band gap through the application of a displacement field normal to the bilayer plane \cite{mccann2006asymmetry,ohta2006controlling,oostinga2008gate}. With a suitable design of top and back gate electrodes, it allows for electrostatic confinement of charge carriers in high quality bilayer graphene devices. These devices use encapsulation in hexagonal boron nitride (hBN) \cite{dean2010boron}, edge contacts \cite{wang2013one} and a graphite back gate \cite{eich2018spin,Ban,doi:10.1021/acs.nanolett.8b01859}, that screens charge impurities trapped in the silicon oxide substrate \cite{zibrov2017tunable}.

In previous experiments, exfoliated graphene flakes have been etched to fabricate graphene nanoribbons \cite{stampfer2009energy,liu2009electrostatic,han2010electron}, single-electron-transistors (SETs) \cite{stampfer2008tunable} and quantum dots \cite{ponomarenko2008chaotic, schnez2009observation} with charge detector \cite{guttinger2012transport,guttinger2008charge,wang2010graphene,fringes2011charge,volk2013probing}. A disadvantage of this fabrication method are charge carrier localizations at the rough sample edges, that can lead to transmission resonances from the tunneling barriers which may dominate the entire device behavior in certain parameter regimes. \cite{bischoff2015localized}.

Here, we use bilayer graphene with its electrostatically induced band gap to fabricate a fully gate-defined device with quantum dots, that are also used as charge detectors. The quality of the bilayer graphene quantum dots and the amplitude of the detection signal matches that of the traditional semiconductors Si and GaAs. A comparison of the charge detection signal in different etched devices with our electrostatically defined device is shown in Table 1. We measure a detection signal that exceeds the best previous results by more than a factor of two. Furthermore, a tunable gate-defined quantum dot is used as charge detector instead of a random resonance in a nano ribbon. The quality of our device and the good control of the charge detector makes it interesting for future investigations of the spin and valley coherence and relaxation rates.

\begin{table}
	\begin{tabular}{l*{6}{c}r}
		\hline
		Publication              & $\Delta G (e^2/h)$ & Fabrication & Detector \\
		\hline
		
		Güttinger et al. (2008) \cite{guttinger2008charge} & $10^{-4}$ & Etched & Resonance in a nanoribbon \\
		Wang et al. (2010) \cite{wang2010graphene}         & $<0.1$ & Etched & SET  \\
		Fringes et al. (2011) \cite{fringes2011charge}   & 0.005 & Etched & Resonance in a nanoribbon\\
		Güttinger et al. (2011) \cite{guttinger2011time}   & 0.05 & Etched & Resonance in a nanoribbon  \\
		Volk et al. (2013) \cite{volk2013probing}   & 0.002 & Etched & Resonance in a nanoribbon  \\
		This work (2019) & 0.2 & Gate-defined & Quantum dot resonance\\
		\hline
		
	\end{tabular} 
	\caption{Comparison of charge detection measurements in graphene in the last years. We compare the charge detection signal as the conductance change $\Delta G$ in the charge detector in devices using different detectors and fabrication methods.}
\end{table}

\section{Characterization}
\begin{figure*}
	\includegraphics[scale=1]{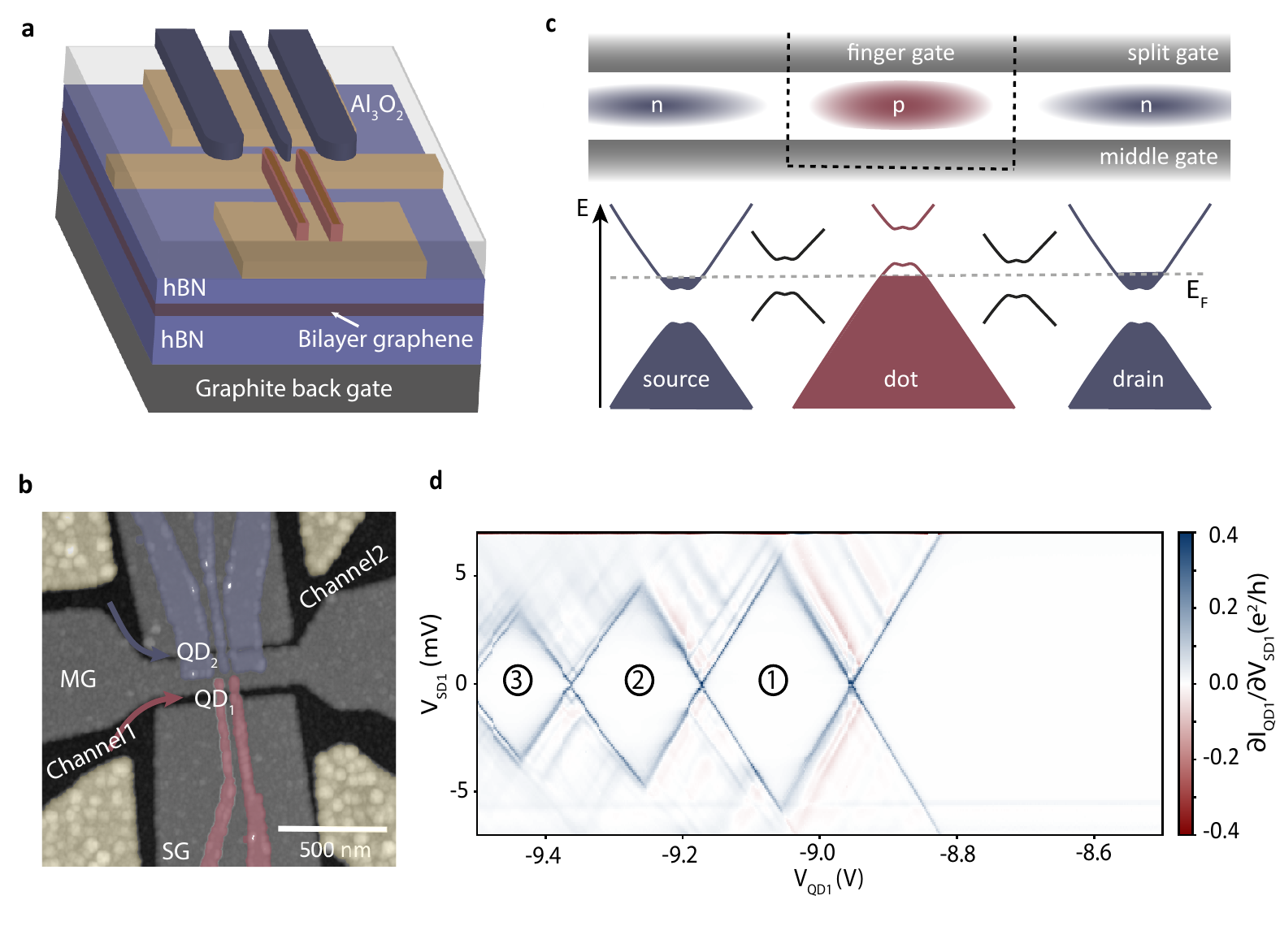}
	\caption{(a) Schematic picture of the different layers in the bilayer graphene device. (b) False-color atomic force microscope (AFM) image of the electrostatically-defined device. By using the split gates SG (gray) in addition to the middle gate MG (gray) two conducting channels (channel 1 and channel 2, black) are created. The finger gates (blue and red) across the channels produce quantum dots and charge detectors in the bilayer graphene. (c) Schematic picture of the band structure at different lateral positions along the current direction (red and blue arrows in (b)) in the channel. The dashed line is the equilibrium electrochemical potential along the direction of current flow. (d) Measurement of Coulomb blockade diamonds of quantum dot 1 (QD1), when the first three holes are charged into the dot.}
	\label{fig1}
\end{figure*}

The van-der-Waals heterostructure has been fabricated using the dry transfer method \cite{dean2010boron}. A schematic of the stack is shown in Figure~\ref{fig1}a. It consists (from bottom to top) of a graphite back gate (grey), a bottom hexagonal boron nitride (hBN) flake (33 nm thick, light blue in Figure~\ref{fig1}a), the bilayer graphene flake (dark red) and a top hBN flake (35 nm thick, light blue). The bilayer graphene flake is electrically contacted using edge contacts \cite{wang2013one} and metal deposition (Cr/Au). In a next step the Cr/Au (5 nm/20 nm for the small structures and 10 nm/60 nm for the connections to the bond pads) split gates and middle gate (SG and MG, grey in Figure~\ref{fig1}b) are deposited on top of the stack. The middle gate is 150 nm wide and separated from the split gates by 100 nm on both sides, forming two channels (black in Figure~\ref{fig1}b). An insulating 30 nm thick $\text{Al}_2\text{O}_3$ layer separates the split gates and the Cr/Au (thickness similar to SGs) finger gates (blue and red in Figure~\ref{fig1}b). The wider finger gates have a width of 120 nm, the narrow one is 20 nm wide. Their lateral separation is 90 nm. A quantum dot can be formed below each of the finger gates (two quantum dots in channel 1 and three dots in channel 2). The atomic force microscope image in Figure~\ref{fig1}b shows the lateral layout of the two top gate layers with split gates (gray), middle gate (gray), finger gates (blue and red) and the Ohmic contacts (yellow).

The graphite back gate and the split gates can be used to (i) open a band gap below the gates, and (ii) tune the Fermi energy into the band gap, which renders these regions insulating. N-type channels with a lithographic width of $100\,\text{nm}$ are formed between the gates (see Figure~\ref{fig1}c blue n-doped regions) by applying a positive voltage to the graphite back gate ($V_{\text{BG}}=3\,\text{V}$) and a negative voltage to the split gates ($V_{\text{SG}}=-3.5\,\text{V}$). The MG is used to separate the channels, where a gate width of $150\,\text{nm}$ is needed to avoid leakage currents between them. Separate source-drain bias voltages $V_{\text{SD}}$ are applied to each channel using pairs of ohmic contacts (Figure~\ref{fig1}b).  

Using one of the finger gates the Fermi-energy in the bilayer graphene region below the finger gate can be tuned into the band gap (pinch-off) or into the p-region (Figure~\ref{fig1}c red), where a QD confining holes is formed. The band structure is sketched in Figure~\ref{fig1}c (lower panel) with source and drain contacts in the n-regions and the hole QD below the finger gate. Between the n- and p-regions the Fermi-energy (at the edges of the finger gate, see Figure~\ref{fig1}c) lies in the band gap, hence natural tunnel barriers for the quantum dot are formed \cite{doi:10.1021/acs.nanolett.8b01859}. Each quantum dot with its sharp Coulomb resonances is also a sensitive detector for single charges in any other QD nearby. The resonances of the QDs are even more sensitive than a QPC, due to the steeper slope of the conductance versus finger gate voltage.

All measurements presented here were performed in a dilution refrigerator with an electronic base temperature of 60 mK in a two-terminal DC setup with a bias voltage applied between source and drain, and the drain contact grounded. The integration time for current measurements was set to 20 ms.

Figure~\ref{fig1}d shows a measurement of Coulomb blockade diamonds of QD1 formed underneath the \textcolor{red}{left} red-colored finger gate in channel 1 (see Figure~\ref{fig1}b). We measure the differential current  $\partial I_{\text{QD1}}/\partial V_{\text{SD1}}$ in channel 1 as a function of the finger gate voltage $V_{\text{QD1}}$ forming QD1 and the DC source-drain bias $V_{\text{SD1}}$. In the Coulomb blockade diamonds we see single holes charging into the quantum dot at $V_{\text{QD1}}=-8.95\,\text{V}, -9.17\,\text{V}, -9.36\,\text{V}$ for $V_{\text{SD}}\approx 0\,\text{V}$. From the Coulomb blockade diamonds a charging energy of about $E_{\text{ch}}=5\,\text{meV}$ and a finger gate lever arm $\alpha=0.02$ can be determined.

\section{Results and discussion}
\begin{figure*}
	\includegraphics[scale=1]{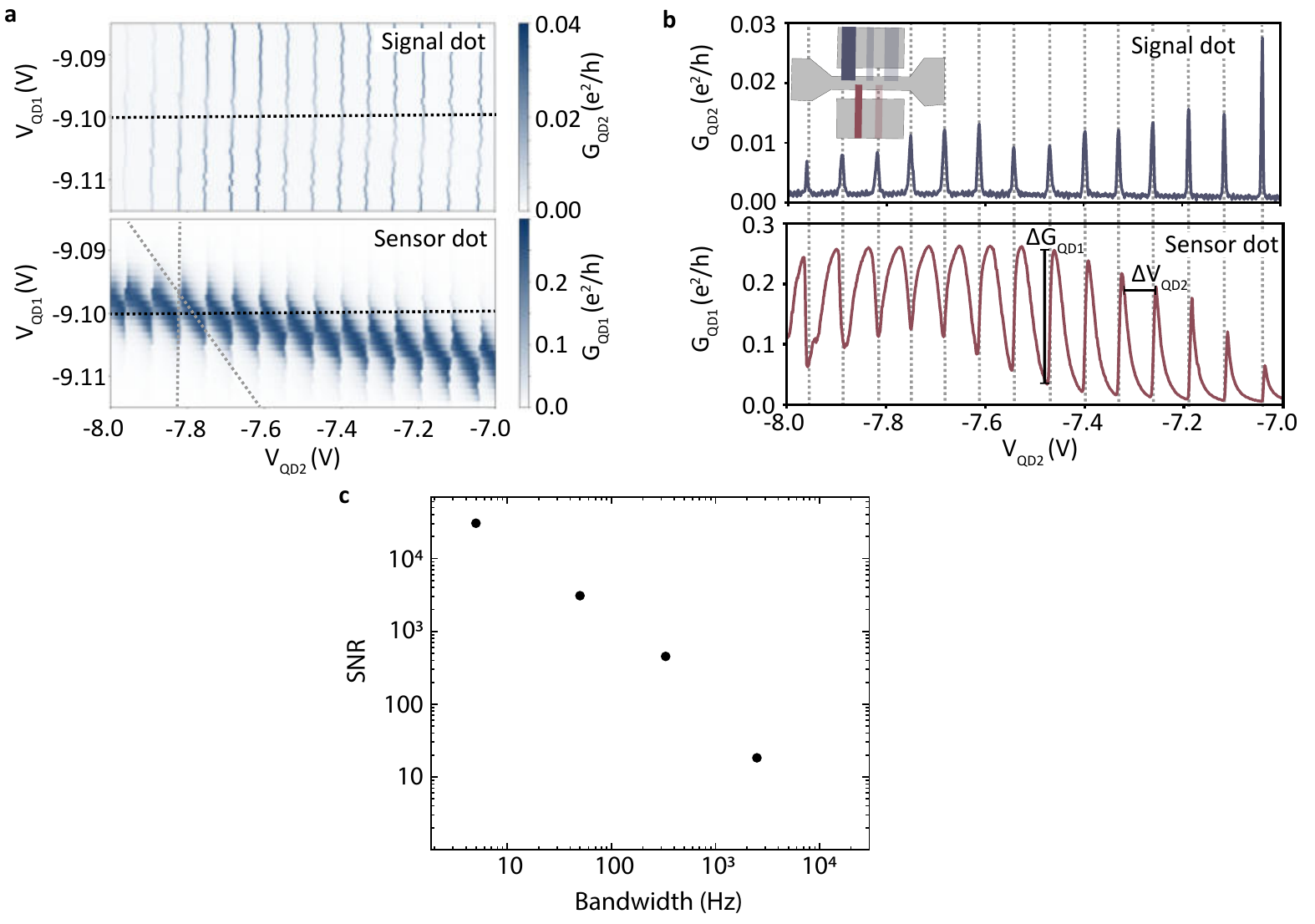}
	\caption{(a) Conductance $G_{\text{QD2}}$ of the signal dot (QD2) (upper panel) and conductance $G_{\text{QD1}}$ of the sensing dot (lower panel) as a function of the finger gate voltages $V_{\text{QD2}}$ and $V_{\text{QD1}}$ for a fixed back gate ($V_{\text{BG}}=3\,\text{V}$) and split gate voltage ($V_{\text{SG}}\approx-3.5\,\text{V}$). Upper panel: The lines spaced with a periodicity of 0.09 V in $V_{\text{QD2}}$ are due to Coulomb blockade resonances. The lower panel shows the simultaneously-acquired measurement of the charge detector conductance $G_{\text{QD1}}$. We observe features aligned with the Coulomb resonances in the upper panel (highlighted with vertical dashed gray line) and tilted lines resulting from the cross capacitance between the sensing dot and $V_{\text{QD2}}$ (highlighted with diagonal dashed gray line). The dashed black lines in the upper and lower panel indicate the line cuts in b, respectively. \textcolor{red}{(c) SNR of the charge detection signal for different measurement system bandwidth.}}
	\label{fig2}
\end{figure*}

For the charge detection experiment shown in Figure~\ref{fig2}, QD1 in channel 1 is used as sensing dot. A source-drain bias of $100\,\upmu\text{V}$ (optimized for the detection signal) is applied across the sensing dot and the finger gate voltage $V_{\text{QD1}}$ is chosen so that sequential tunneling through QD1 is possible and a current is measured. At this finger gate voltage the quantum dot is a sensitive detector for changes of the charge configuration in its environment. A small change in the electrostatic environment of the dot leads to a shift of the Coulomb resonance in energy (or equivalently, in finger gate voltage $V_{\text{QD2}}$) and therefore to a change in the conductance $G_{\text{QD1}}$ through channel 1. 

A second p-type quantum dot (signal dot QD2) is formed below the left blue gate in channel 2 (Figure~\ref{fig1}a), by tuning the finger gate voltage $V_{\text{QD2}}$. The linear conductance $G_{\text{QD2}}$ through channel 2 is shown in Figure~\ref{fig2}a upper panel, where the gate voltages $V_{\text{QD1}}$ and $V_{\text{QD2}}$ are changed and a source-drain voltage of $V_{\text{SD}}=100\,\upmu\text{V}$ across the signal dot is applied. By changing $V_{\text{QD2}}$ we scan across one Coulomb resonance of QD1 (detector dot) and measure the conductance $G_{\text{QD1}}$ in channel 1 and $G_{\text{QD2}}$ channel 2 at the same time. The upper panel in Figure~\ref{fig2}a shows regularly spaced Coulomb resonances in the conductance of the signal dot, which slightly shift by changing $V_{\text{QD1}}$. 

These resonances are observed through charge detection in the lower panel of Figure~\ref{fig2}a. A line-cut in $V_{\text{QD1}}$ direction shows a Coulomb resonance of the sensing dot (QD1), that shifts to more negative $V_{QD1}$, when $V_{QD2}$ is increased. The diagonal shift of the sensing dots resonance (marked by diagonal dashed line) is due to the cross capacitance between the sensing dot and the finger gate voltage defining the signal dot. From this shift a ratio between the lever arm of the blue gate ($V_{\text{QD2}}$) on the sensor dot and the red gate ($V_{\text{QD1}}$) on the sensor dot $\alpha_{V_{\text{QD2}}-\text{QD1}}/\alpha_{\text{$V_{\text{QD1}}$-QD1}}=0.07$ is calculated. 

We identify single charging events in the signal dot as abrupt shifts of the conductance resonance in the sensing dot (marked with vertical dashed line), aligned with the Coulomb resonances of the signal dot (see vertical dashed lines). Line cuts from Figure~\ref{fig2}a at $V_{\text{QD1}}=9.1\,\text{V}$ are shown in Figure~\ref{fig2}b. Regularly spaced conductance resonances are observed in Figure~\ref{fig2}b (upper panel), when sequential tunneling through the signal dot is possible. 

The corresponding conductance $G_{\text{QD1}}$ measured simultaneously in the detector channel is shown in the lower panel in Figure~\ref{fig2}b. We observe a broadened resonance with its maximum at $V_{\text{QD2}}=-7.75\,\text{V}$ and a width of $1\,\text{V}$ with step-like features on top. The resonance is broader in $V_{\text{QD2}}$ than in $V_{\text{QD1}}$, due to the much smaller lever arm of the blue finger gate on the sensing dot as compared to the red finger gate. The conductance steps (marked with vertical dashed lines) are related to a shift $\Delta V_{\text{QD2}}$ in the  resonance of the sensing dot with respect to the signal dot's voltage $V_{\text{QD2}}$. 
 
From an analysis of the charging events an average shift in $\Delta V_{\text{QD2}}=61\,\text{mV}$ (see Figure~\ref{fig2}b) is observed. The conductance in the detector channel $\Delta G_{\text{QD1}}=0.2\,e^2/h$ (see Figure~\ref{fig2}b) changes by up to 77\% for a single charging event. This change in conductance is compared to other devices in Graphene in Table~1. Furthermore, we evaluate the signal to noise ratio (SNR) of $\Delta I_{Step}/<I_{Noise}>\approx 3000$ at a measurement bandwidth of 50 Hz. $<I_{Noise}>$ is the variance of the detector current and $\Delta I_{Step}$ the current change in the detector due to one charging event. In Figure \ref{fig2}c the SNR of the charge detection signal is shown for different measurement bandwidth. From these measurements we expect a maximal bandwidth of 10 kHz for our measurement, which is higher than in etched graphene devices \cite{guttinger2011time}. The tunneling rates in the presented device are higher than the maximal measurement bandwidth, so that time-resolved measurements of electron tunneling is not possible in this device.

\begin{figure}
	\includegraphics[scale=1]{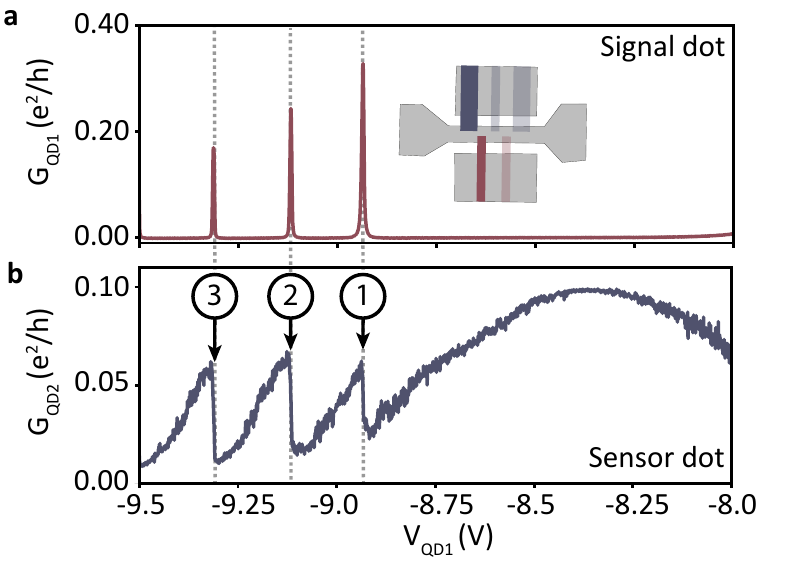}
	\caption{(a) Conductance $G_{\text{QD1}}$ of the signal dot and (b) conductance $G_{\text{QD2}}$ of the sensing dot as a function of the finger gate voltage $V_{\text{QD1}}$ for a fixed back gate ($V_{\text{BG}}=3\,\text{V}$) and split gate voltage ($V_{\text{SG}}=-3.5\,\text{V}$). (a) The first three Coulomb resonance of the signal dot are measured directly in the current through the dot and (b) in the charge detector. At $V_{\text{QD1}}=-8.9\,\text{V}$ the first Coulomb resonance and also the first step in the detector signal is measured, proofing the charging of the QD with the first hole.}
	\label{fig3}
\end{figure}

In the following, we exchange the role of the two dots, to confirm that we are able to fully deplete the quantum dot and fill it with individual holes. The quantum dot in channel 2 (QD2) will be used as the sensing dot. The conductances through the signal dot in channel 1 (QD1) is shown in Figure~\ref{fig3}a. For QD1, we can clearly see a first Coulomb resonance (marked by \textcircled{1}) in the conductance of the channel 1 at $-8.9\,\text{V}$. The conductances $G_{\text{QD1}}$ and $G_{\text{QD2}}$ of both channels are shown in Figure~\ref{fig3}a and b, while the gate voltage of the signal dot (QD1) is changed and a source-drain bias of $V_{\text{SD}}=100\,\upmu\text{V}$ is applied to both channels. In the conductance of the detector $G_{\text{QD2}}$ a first step is also observed at about $-8.9\,\text{V}$. 

\begin{figure*}
	\includegraphics[scale=1]{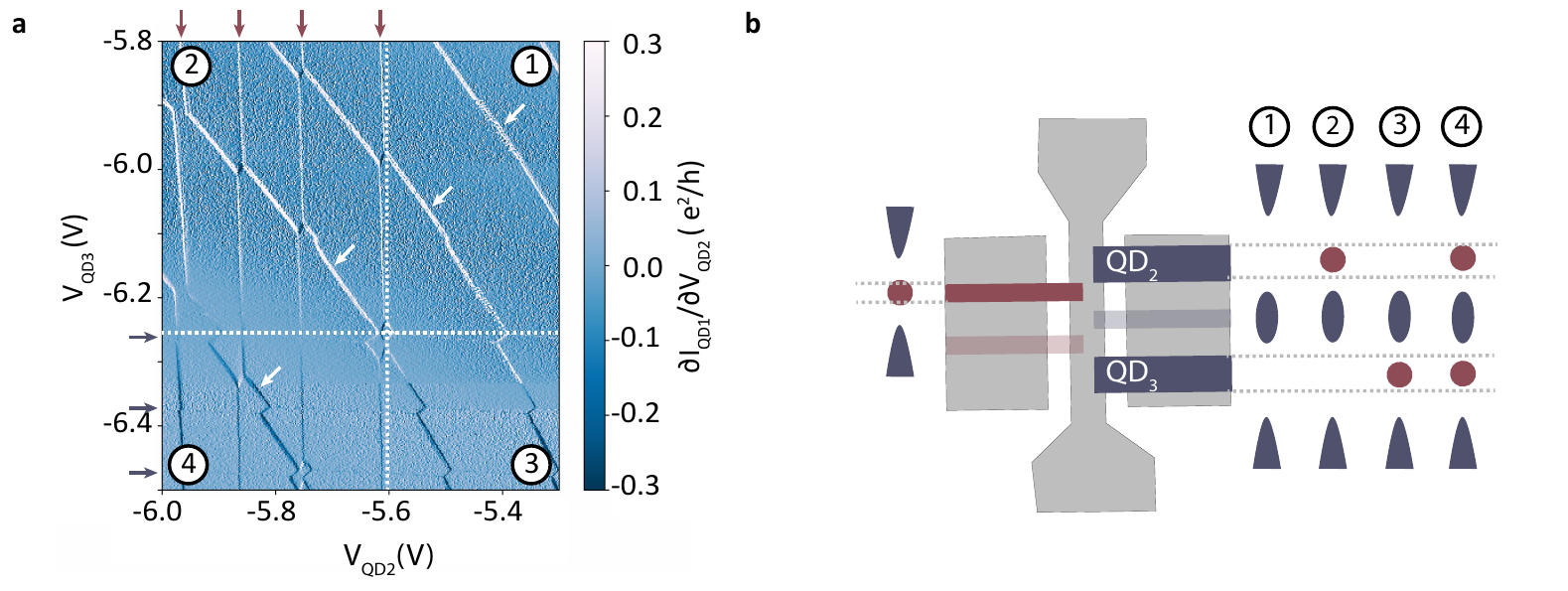}
	\caption{(a) Differential conductance $\partial I_{\text{QD1}}/\partial V_{\text{QD2}}$ in the detection channel, when multi dots are formed in channel 2. The white dotted lines indicate the boundaries of four different regions. In region \textcircled{1}, a single electron dot is formed between two barriers. In regions \textcircled{2} and \textcircled{3}, an electron-hole double dot is formed between the gates and below one of the gates, respectively. In region \textcircled{4}, a triple dot is measured. The three dots are formed below each of the gates and between them. (b) Schematic picture of the device showing the different dot configurations in (a). Red dots and blue dots show the QDs formed below the gates (hole dots) and between them (electron dots), respectively.}
	\label{fig4}
\end{figure*}

Furthermore, QDs with tunable tunneling barriers and multiple-dots can be formed in channel 2 using the two broader finger gates (blue in Figure~\ref{fig4}b).\cite{doi:10.1021/acs.nanolett.8b01859} Figure~\ref{fig4}a shows the differential conductance of the charge detector $\partial I_{\text{QD1}}/\partial V_{\text{QD2}}$ (the conductance change through the signal dot is lower than our measurement resolution) as function of the gate voltages $V_{\text{QD2}}$ and $V_{\text{QD3}}$ that were applied to the two outer finger gates in channel 2. Three distinct sets of resonances are observed vertical (marked with red arrows), horizontal (marked with blue arrows) and diagonal (marked with white arrows) resonances in Figure~\ref{fig4}a). The measurement can be divided into four quadrants (\textcircled{1} to \textcircled{4}), separated by the white dashed lines. The corresponding sketch of the charge carrier distributions along channel 2 is shown in Figure~\ref{fig4}b. In the first quadrant of Figure~\ref{fig4}a we observe diagonal resonances only, that belong to a single electron dot formed between the two outer gates in channel 2 (dark blue in Figure~\ref{fig1}a and Figure~\ref{fig4}b). The two finger gates are tuned close to the charge neutrality point, thus creating tunneling barriers between the source and drain contacts (see Figure~\ref{fig4}b \textcircled{1}). Both gates have the same lever arm on the resonances of this dot, leading to diagonal resonances with a slope of about -1 in this measurement. The tunneling barriers of this dot are tunable, which allows for changing the tunneling rates through the dot.

By decreasing the voltages of the finger gates $V_{\text{QD2}}$ and $V_{\text{QD3}}$ further, we can form double- and triple-dots in channel 2 (sketches of the charge carrier density in the channel are shown in Figure~\ref{fig4}b \textcircled{2} to \textcircled{4}) similar as demonstrated in Ref.~\cite{doi:10.1021/acs.nanolett.8b01859}. Charge occupation of these multidots, is also detected using QD1 as the charge detector.
On the one hand, the gate voltage $V_{\text{QD2}}$ on the left outer finger gate in channel 2 is more negative and a hole dot forms below the gate. Hence, additional vertical resonances (marked with red arrows) are observed in region \textcircled{2} at $V_{\text{QD2}}=-5.61\,\text{V}, -5.76\,\text{V}, -5.87\,\text{V}$ and $-5.97\,\text{V}$ (see schematic picture \textcircled{2} in Figure~\ref{fig4}b). For this situation of an electron-hole double-dot, we observe the typical honeycomb pattern using the charge detector.

On the other hand, we decrease the voltage on finger gate 3 ($V_{\text{QD3}}$) such that a hole QD is formed below this gate. We observe horizontal resonances in region \textcircled{3} in Figure~\ref{fig4}a ($V_{\text{QD3}}=-6.26\,\text{V}, -6.38\,\text{V}$ and $-6.47\,\text{V}$, marked with blue arrows), which are not influenced by $V_{\text{QD2}}$. In region \textcircled{4} in Figure~\ref{fig4}a, three dots are formed- one p-type QD below each of the gates and an n-type dot between them. This leads to diagonal resonances for the tunneling through the electron dot in the middle, horizontal resonances for tunneling through QD3 and vertical line for tunneling through QD2.

The three dots measured in Figure~\ref{fig4}a have different distances to the charge detector. Hence, charging the different dots leads to different energy shifts of the sensing dots resonance. The distance between the signal dot and the sensing dot changes from $260\,\text{nm}$ (dot between the gates, blue in Figure~\ref{fig4}b) to $315\,\text{nm}$ (dot below left gate in Figure~\ref{fig4}b). The energy shift of the detector resonance is evaluated from the detection signal using the shift in gate voltage and the relative lever arms $\alpha$,  $\alpha_{\text{FG2-QD1}/\alpha_{\text{FG1-QD1}}}$ and  $\alpha_{\text{FG3-QD1}/\alpha_{\text{FG1-QD1}}}=0.01$. It decreases with the distance between sensing and signal dot from $50\,\upmu \text{eV}$ to $5\,\upmu \text{eV}$ in agreement with a Coulomb screening model. Our sample design allows for a minimal distance of $150\,\text{nm}$ between the sensing dot and the signal dot, which leads to an energy shift of $6\,\text{meV}$ which is of the same order as the charging energy of the sensing dot.

In conclusion, we presented an electrostatically-defined device that allows us to detect single charge carriers in bilayer graphene quantum dots. Using conductance resonances in the Coulomb-blockade regime of a second quantum dot as a sensitive detector, we reached a maximum relative conductance change of 77\% for charge detection. Our measurements show that a width of $150\,\text{nm}$ for the MG is sufficient to avoid leakage between the sensing and signal dot and to reach a high signal-to-noise ratio in the detector. Using this device, we were able to show complete depletion of one of the quantum dots. Furthermore, we were able to observe the changes in the charge state of a quantum dot with tunable tunneling barriers, and of a multi-dot system. In the multi-dot regime, the charge detection enables us to determine the number of charge carriers in each of the dots. Our experiments demonstrate a device that is needed as the starting point for time-resolved measurements in graphene quantum dots, which may allow us to investigate the spin-lifetime in graphene \cite{elzerman2004single}. 

\section{Acknowledgments}
We thank Peter Märki, Erwin Studer, as well as the FIRST staff for their technical support. We also acknowledge financial support from the European Graphene Flagship, the Swiss National Science Foundation via NCCR Quantum Science and Technology, the EU Spin-Nano RTN network and ETH Zurich via the ETH fellowship program. Growth of hexagonal boron nitride crystals was supported by the Elemental Strategy Initiative conducted by the MEXT, Japan and the CREST (JPMJCR15F3), JST.

\providecommand{\latin}[1]{#1}
\makeatletter
\providecommand{\doi}
{\begingroup\let\do\@makeother\dospecials
	\catcode`\{=1 \catcode`\}=2 \doi@aux}
\providecommand{\doi@aux}[1]{\endgroup\texttt{#1}}
\makeatother
\providecommand*\mcitethebibliography{\thebibliography}
\csname @ifundefined\endcsname{endmcitethebibliography}
{\let\endmcitethebibliography\endthebibliography}{}

\newpage
\begin{figure*}
	\includegraphics[scale=1]{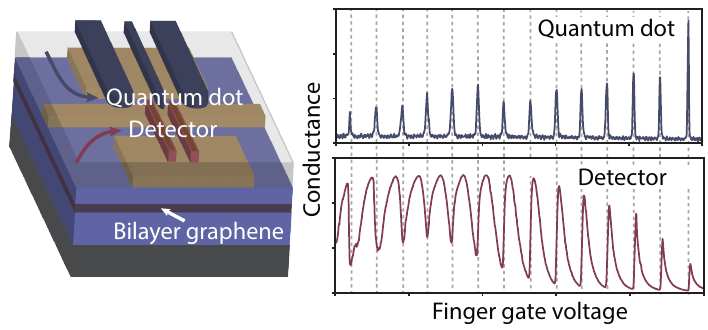}

\end{figure*}

\end{document}